\documentclass[preprint, superscriptaddress, aps, prb, showpacs]{revtex4-2}

\usepackage{chemformula} 
\usepackage[T1]{fontenc} 
\usepackage{orcidlink}
\usepackage{multirow, tabularx, booktabs}
\usepackage{color}
\usepackage{upgreek}
\usepackage{graphicx}
\usepackage{makecell}
\usepackage{amsmath}
\usepackage{float}
\usepackage{adjustbox}
\usepackage{array}
\usepackage{booktabs}
\usepackage{ragged2e}
\usepackage{indentfirst}
\newcolumntype{Y}{>{\RaggedRight\arraybackslash}X}

\usepackage{bm,amsmath,amssymb}
\usepackage[version=4]{mhchem}
\hypersetup{
	colorlinks=true,
	linkcolor=blue,
	citecolor=blue,
	urlcolor=blue
}

\begin{document}

\title{Distinguish the Orientation of Sliding Ferroelectricity by Second-Harmonic Generation}

\author{Fengfeng Ye}
\altaffiliation{These authors contributed equally to this work}
\affiliation{Frontier Institute of Science and Technology, State Key Laboratory of Electrical Insulation and Power Equipment, Xi’an Jiaotong University, Xi’an 710049, People's Republic of China}

\author{Qiankun Li}
\altaffiliation{These authors contributed equally to this work}
\affiliation{School of Physical Science and Technology, Jiangsu Key Laboratory of Frontier Material Physics and Devices, Soochow University, Suzhou, 215006, China}

\author{Zhuocheng Lu}
\altaffiliation{These authors contributed equally to this work}
\affiliation{Center for Quantum Matter, School of Physics, Zhejiang University, Zhejiang 310058, China}

\author{Xinfeng Chen}
\affiliation{Frontier Institute of Science and Technology, State Key Laboratory of Electrical Insulation and Power Equipment, Xi’an Jiaotong University, Xi’an 710049, People's Republic of China}

\author{Yang Li}
\affiliation{School of Materials Science and Engineering, Harbin Institute of Technology, Harbin 150001, China}

\author{Hua Wang}
\email{daodaohw@zju.edu.cn}
\affiliation{Center for Quantum Matter, School of Physics, Zhejiang University, Zhejiang 310058, China}

\author{Lu You}
\email{lyou@suda.edu.cn}
\affiliation{School of Physical Science and Technology, Jiangsu Key Laboratory of Frontier Material Physics and Devices, Soochow University, Suzhou, 215006, China}

\author{Gaoyang Gou}
\email{gougaoyang@mail.xjtu.edu.cn}
\affiliation{Frontier Institute of Science and Technology, State Key Laboratory of Electrical Insulation and Power Equipment, Xi’an Jiaotong University, Xi’an 710049, People's Republic of China}

\date{\today}

\begin{abstract}

As the emerging ferroelectric (FE) materials, the ultrathin two-dimensional (2D) sliding ferroelectrics without phase-matching bottleneck, usually exhibit the pronounced second harmonic generation (SHG) responses. Despite the structural polarity of sliding ferroelectrics can be precisely detected via SHG characterizations, distinguishing the orientations of sliding ferroelectricity based on SHG responses has rarely been realized, as SHG intensities for upward and downward polarization states are supposed to be same. In current work, combining computational simulations and experimental characterizations, the orientation of sliding ferroelectricity is demonstrated to be readily distinguishable via SHG responses in 2D SnP$_2$S$_6$ (SnP$_2$Se$_6$), a new sliding FE material. Specifically, owing to the unique symmetry operation within FE-SnP$_2$S$_6$ (SnP$_2$Se$_6$), the intersection between $\chi_{xxx}$ and $\chi_{yyy}$ SHG susceptibility coefficients with opposite signs leads to the effective rotation of SHG polar directions upon switching of sliding ferroelectricity. Moreover, the remarkable dependence of SHG polar directions on the orientation of sliding ferroelectricity is further validated by experimental characterizations performed on SnP$_2$S$_6$ crystal in a single FE domain structural form. This work opens up the avenue for in-situ detecting the ferroelectricity orientation of 2D sliding ferroelectrics based on SHG nonlinear optical responses, and also demonstrates the controllable optical nonlinearly for new “slidetronics” applications.

\end{abstract}

\maketitle
\clearpage

Second-order nonlinear optical (NLO) properties, such as SHG effect, are indispensable for optoelectronic and photonic applications including high-power laser, frequency conversion and photon modulation\cite{000714972500014,000849306100001}. Especially, NLO crystals with large SHG susceptibility and easy phase-matchability are highly desirable for the efficient NLO conversion. 2D van der Waals (vdW) layered materials with relaxed phase-matching requirement can exhibit the SHG efficiency even superior than the bulk NLO crystal such as Perovskite LiNbO$_3$\cite{000955711300008}. In fact, large SHG susceptibility have been detected in transition metal dichalcogenides (e.g. MoS$_2$\cite{000319056400002,000385428300002,000430642500012}, MoSe$_2$\cite{001327106200001}, MoTe$_2$\cite{000443548600002,000413697800036}, WS$_2$\cite{000874860200001,000711790600111}, WSe$_2$\cite{000350976700028,000391449600003}, ReS$_2$\cite{000445322300007}) monolayers, h-BN monolayer\cite{001357821000001,000407540300071} and 2D displacive FE materials including CuInP$_2$S$_6$\cite{000903327500001} and $\alpha$-In$_2$Se$_3$\cite{000772917900002,000456421000002}. From the structural point of view, the absence of spatial inversion symmetry in 2D materials manifests as a fundamental requirement for nonzero SHG responses. Therefore, SHG effect is an effective optical approach for precisely detecting the structural polarity of 2D materials.

Thanks to the rising of sliding ferroelectricity\cite{000404808000122}, the noncentrosymmetric 2D layered structures can be realized by combining the parent centrosymmetric monolayers in a specific interlayer stacking sequence. Accumulation of interlayer dipoles can generate the nonzero out-of-plane ferroelectricity, whose orientations can be further switched via sliding of individual monolayers along the in-plane directions\cite{000732715700052}. Coupling of sliding ferroelectricity with intrinsic properties of 2D materials further give arise to the rich functionality, including the electric controllable nonlinear Hall effect\cite{000550580800009}, Berry curvature memory\cite{000544166500001}, bulk photovoltaic\cite{001380711000001}, ferro-valley\cite{001457028700005} and spintronics\cite{001180085100001} effects. Moreover, the upward and downward polarization ($\pm$$P$) states also correspond to the logical On/Off status for minimized electronic devices based on sliding ferroelectrics\cite{001336336000017}. Therefore, orientation of sliding ferroelectricity manifests as a unique degree of freedom for performance engineering of 2D sliding ferroelectrics. In this context, in-situ and high-precision detecting the orientation of sliding ferroelectricty is highly desirable in the field of 2D “slidetronics”\cite{001360291200001}. However, for sliding ferroelectrics with the subtle polarization magnitudes and small cohesive electric field\cite{000981076000001}, most traditional characterizing approach becomes incapable to determine their polarization orientations. Detecting the subtle structural polarity of sliding ferroelectrics based on SHG characterizations have been extensively performed in experiment\cite{000754187400008,001158421900001}, but orientation of sliding ferroelectricity are hardly distinguishable based on SHG responses. As experimentally detectable SHG intensity $I$ $\propto$ $|\chi|^2$, $\pm$$P$ states for the nonmagnetic sliding ferroelectrics have exactly same crystallographic symmetry and SHG susceptibility magnitudes $|\chi|$\cite{001206118800001}. Therefore, even $\chi$ coefficient changes its sign, the overall SHG responses (including magnitudes and polarimetry) of sliding ferroelectrics remain invariant upon polarization switching.

In current work, based on crystal symmetry analysis, we reveal that above limitations indeed apply for most known sliding ferroelectrics, whose polarization orientations are not distinguishable via SHG characterizations. As the rare exception, 2D layered SnP$_2$S$_6$ and SnP$_2$Se$_6$ with unique crystallographic symmetry have the multiple SHG suscepltibility coefficients ($\chi_{xxx}$ and $\chi_{yyy}$) contributing to the experimental SHG responses. Our first-principles calculations demonstrate 2D vertically polarized SnP$_2$S$_6$ (SnP$_2$Se$_6$) multilayers as sliding ferroelectrics, whose polarizations can be reversed via sliding of individual monolayers. Moreover, based on SHG susceptiblity simulations, it is predicted that $\chi_{xxx}$ coefficient changes its sign while $\chi_{yyy}$ remains invariant after switching of polarization. Incident by the linearly polarized light with a characteristic photo frequency, the ideal intersections between $\chi_{xxx}$ and $\chi_{yyy}$ with opposite signs and nearly identical magnitudes can lead to almost 30$^\circ$ rotation SHG polar directions upon polarization switching. Based on theoretical predictions, the remarkable dependence of SHG polar directions on the orientation of sliding ferroelectricity is detected by the experimental SHG characterizations performed on SnP$_2$S$_6$ sample in single domain structural form.

As the experimentally synthesized 2D layered materials, tin hypodithiophosphates SnP$_2$S$_6$\cite{A1995QY89900007} and its structural analogue SnP$_2$Se$_6$\cite{000980689000021} both adopt the polar rhombohedral phases in $R3$ space group (detailed crystallographic parameters tabulated in TABLE S1 of Supporting Information). Shown in Figure 1a is the crystal structure of SnP$_2$S$_6$ in hexagonal setting, bulk SnP$_2$S$_6$ contains three rhombohedral-stacked monolayers in $ABC$ stacking sequence. Within each monolayer, 2/3 sulfur octahedral sites are occupied by Sn$^{4+}$ cations and P-P dimers, while the other 1/3 octahedral sites remain unoccupied, forming SnS$_6$ and P$_2$S$_6$ octahedrons in edge-sharing connectivity. Owing to the broken inversion symmetry along the out-of-plane direction, SnP$_2$S$_6$ (SnP$_2$Se$_6$) is a 2D vertically polarized material. On the one hand, structural polarity of SnP$_2$S$_6$ could originate from the vertical polar displacement of Sn cations relative to sulfur octahedral centers, similar to 2D FE CuInP$_2$S$_6$ with displacive ferroelectricity\cite{000381525200001,000544063600028}. On the other hand, rhombohedral $ABC$ stacking within SnP$_2$S$_6$ can also produce the net interlayer dipoles via relative sliding between adjacent monolayers, as demonstrated in sliding ferroelectrics such as 3R-MoS$_2$\cite{001336336000012,001336336000017} and 3R-BN\cite{001262373000024}.

\begin{figure}[h]
	\centering
	\includegraphics[width=1\linewidth]{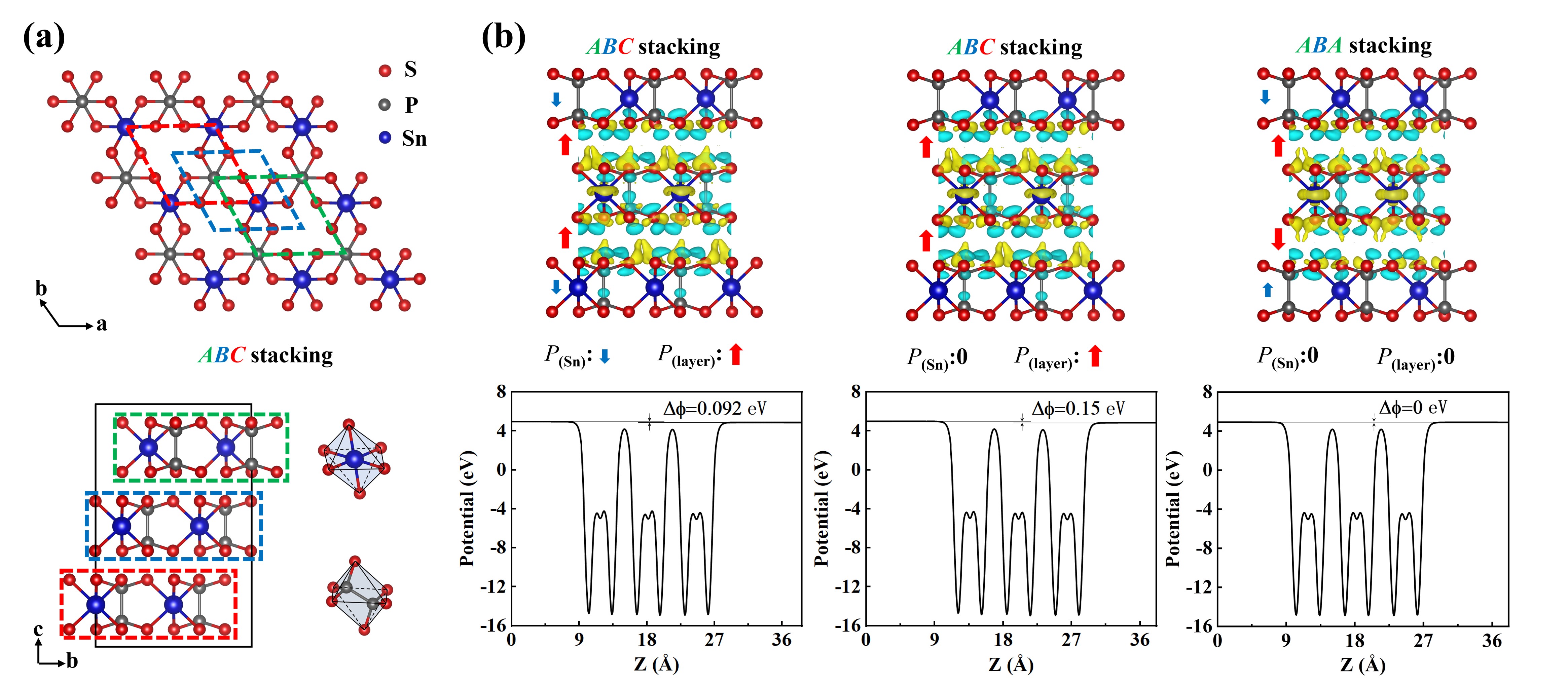}\vspace{-3pt}
	\caption{(a) Atomic structure for vdW layered SnP$_2$S$_6$, composed of three monolayers in $ABC$ stacking sequence. Both top and side views are provided. (b) Top: simulated interlayer charge density of raw SnP$_2$S$_6$ trilayer “exfoliated” from the bulk phase, SnP$_2$S$_6$ trilayer in $ABC$ stacking but without cation polar displacement and trilayer with cation polar displacement and $ABA$ stacking sequence. Charge accumulation and depletion of the as-formed interlayer dipoles are indicated by yellow and cyan isosurface plots with the isovalue of $\pm 1.8 \times 10^{-4} \, \text{e}/\text{Å}^3$, respectively. Bottom: the planar-averaged electrostatic potential profiles for SnP$_2$S$_6$ trilayers, where the out-of-plane polarizations create the non-zero potential offsets between two ends.}
	\label{Figure 1}
\end{figure}

In order to reveal the macroscopic origin responsible for out-of-plane polarization, structural analysis has been conducted. Both SnP$_2$S$_6$ and SnP$_2$Se$_6$ are vdW layered materials with weak interlayer binding energies ($E$\textsubscript{b} = $-18.29$ and $-21.02$ meV/Å$^2$ for SnP$_2$S$_6$ and SnP$_2$Se$_6$), we can therefore “exfoliate” SnP$_2$S$_6$ trilayer from the bulk phase. As shown in Figure 1b, out-of-plane polarization persists in the “exfoliated” 2D SnP$_2$S$_6$ trilayer, which manifests as the non-zero electrostatic potential offset ($\Delta \upphi$ = 0.092 eV) between two ends of trilayer. After eliminating the vertical Sn cation polar displacement, SnP$_2$S$_6$ trilayer composed of three rhombohedral-stacked centrosymmetric monolayers, exhibits the even larger potential offset ($\Delta \upphi$ = 0.15 eV). In fact, Sn polar displacement produces the displacive $P_{(Sn)}$ opposite to overall polarization (see TABLE S2 for details). Out-of-plane polarization within SnP$_2$S$_6$ is indeed contributed by the interlayer dipoles associated with $ABC$ stacking, which are illustrated by differential charge density of trilayer relative to individual centrosymmetric monolayers shown in Figure 1b. Moreover, alternating the stacking sequence of SnP$_2$S$_6$ has a significant consequence on its out-of-plane polarization. Specifically, for SnP$_2$S$_6$ trilayer with $ABA$ stacking, the net interlayer dipole between adjacent monolayer remains, but overall polarization completely cancels out owing to the opposite interlayer dipole arrangement. Based on the analysis above, out-of-plane polarization of SnP$_2$S$_6$ is a direct consequence of interlayer sliding associated with $ABC$ stacking, rather than net polar displacement of Sn cations. Therefore, SnP$_2$S$_6$ (SnP$_2$Se$_6$) is a potential 2D sliding FE material, similar to 3R-MoS$_2$ and 3R-BN.

As a potential 2D sliding FE material, nature of siding ferroelectricity requires that out-of-plane polarization of SnP$_2$S$_6$ (SnP$_2$Se$_6$) is reversible via sliding of individual monolayers. To this end, we first examine the interlayer sliding energy landscape by simulating the potential energy surface (PES) associated with relative sliding between two centrosymmetric SnP$_2$S$_6$ monolayers. As shown in Figure 2a, large red circles in PES correspond to the stable SnP$_2$S$_6$ bilayer configurations in $AB$ or $BA$ stacking. Transition from one stable state to other neighboring stable configurations can be realized by sliding one monolayer along the crystallographic $a(b)$ axis or diagonal [$\overline1$10] direction. Clearly, the optimal interlayer sliding pathway is along [$\overline1$10] direction (indicated by red arrow), as the system will encounter the lower energy barrier without passing through the unfavorable $AA$-stacked structure.

\begin{figure}[!htb]
	\centering
	\includegraphics[width=1\linewidth]{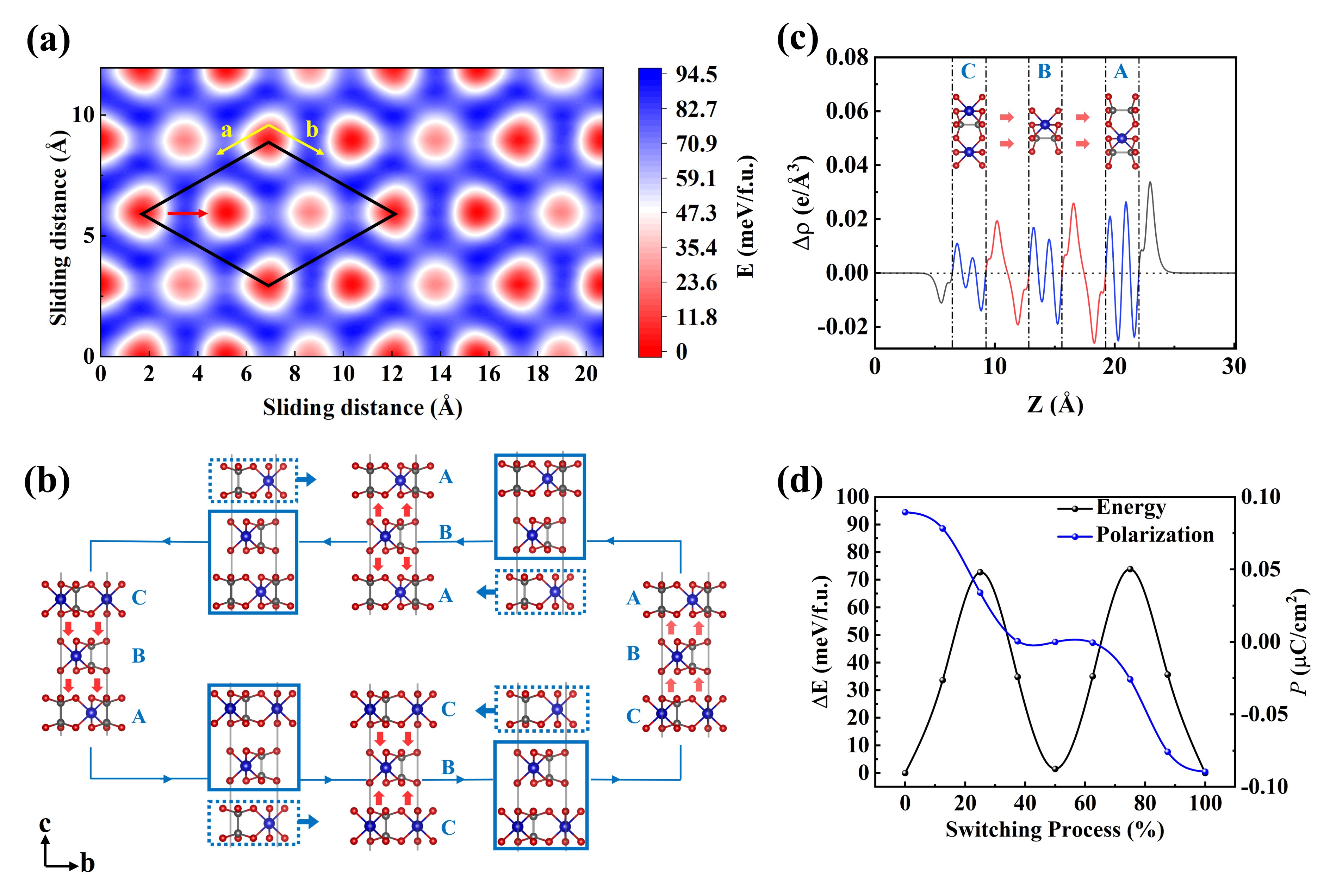}\vspace{-3pt}
	\caption{(a) Our simulated PES associated with the sliding between two SnP$_2$S$_6$ monolayers. The unit cell of SnP$_2$S$_6$ is marked by a black diamond, the red arrow indicates the direction for optimal interlayer sliding pathway along [$\overline{1}$10] direction, and the yellow arrow is the lattice vector. (b) The optimal polarization switching pathway via consecutive sliding of individual monolayers between trilayer structures with $ABC$ and $CBA$ stacking sequences. The solid and dashed rectangles indicate the stationary and sliding monolayers, respectively. Solid blue arrows mark the sliding directions of monolayers. The interlayer dipoles between adjacent monolayers are represented by red arrows. (c) The simulated planar-averaged screening charge $\Delta \rho$ along the out-of-plane direction for +$P$ state of SnP$_2$S$_6$ trilayer, where $\Delta \rho$ from interlayer and intralayer areas are in red and blue colors, respectively. (d) Along the optimal interlayer slidng pathway, variation of system energy and polarization magnitude $P$ during the polarization switching between $\pm$$P$ states  of SnP$_{2}$S$_{6}$ trilayer.}
	\label{Figure 2}
\end{figure}

Using SnP$_2$S$_6$ trilayer as a prototypical case, we further illustrate how its polarization can be switched by the relative sliding of adjacent monolayers. As shown in Figure 2b, $\pm$$P$ states of SnP$_2$S$_6$ trilayer refer to structures with reversed interlayer stacking sequences ($ABC$ $vs$ $CBA$). Switching out-of-plane polarization is equivalent to transition from $ABC$ to $CBA$-stacked trilayer structure. Typically, along the interlayer sliding induced polarization switching pathway, top two monolayers of $ABC$-stacked SnP$_2$S$_6$ trilayer remain stationary, while the bottom monolayer slides along [$\overline1$10] direction. After crossing a saddle point state, the system reaches an intermediate phase in $ABA$ stacking. Polarization switching process continues by sliding of top monolayer along [$\overline1$10] direction. After crossing the other saddle point, SnP$_2$S$_6$ trilayer reaches $CBA$-stacked configuration with opposite polarization. Meanwhile, $CBA$-stacked trilayer transforms into $ABC$-stacked structure along the symmetry equivalent pathway with $CBC$-stacked trilayer as the intermediate phase. 

The polarization magnitude $P$ for all configurations along the sliding pathway can be simulated via the integration of planar averaged screening charge density\cite{001180085100001} ($\Delta \rho$, charge density difference relative to the centrosymmetric reference). Specifically, Figure 2c displays $\Delta \rho$ along the out-of-plane direction for $+P$ state of SnP$_2$S$_6$ trilayer. Integrating $\Delta \rho$ over the effective thickness $L$ yields the net dipole moment $D$. Out-of-plane $P = D/L = 0.09$ $\mathrm{\mu C/cm^2}$ is then predicted for $+P$ state of SnP$_2$S$_6$ trilayer. After recording the system energy and $P$ for each configuration along the pathway, polarization and energy curves associated with polarization switching via interlayer sliding are obtained. As displayed by Figure 2d, the intermediate $ABA$ or $CBC$ trilayer configuration corresponds to paraelectric (PE) phase with $P$ = 0. SnP$_2$S$_6$ trilayer encounters a mediate polarization switching barrier ($\Delta$$E$ = 72.8 meV/f.u.). Beside the consecutive sliding of individual monolayer, $P$ of SnP$_2$S$_6$ trilayer can also be switched via simultaneous sliding two of three monolayers, but encountering the larger $\Delta$$E$ (see Figure S1 of Supporting Information for details). Moreover, SnP$_2$Se$_6$ trilayer can also switch its polarization along the optimal interlayer sliding pathway isostructural to SnP$_2$S$_6$ (Figure S2 and Figure S3). Therefore, combining the stacking sequence dependent out-of-plane polarization and interlayer sliding induced polarization switching, SnP$_2$S$_6$ and SnP$_2$Se$_6$ are demonstrated as a new group of 2D sliding ferroelectrics.

Owing to the breaking of inversion symmetry and relaxed phase-matching requirement, strong SHG responses have been detected in 2D SnP$_2$S$_6$ and SnP$_2$Se$_6$ nanoflakes\cite{000804570900030,000693361000008,000980689000021}. Figure 3a illustrates the typical SHG polarimetry geometry, where the frequency doubling effect occurs when incident light interacts with FE SnP$_2$S$_6$ (SnP$_2$Se$_6$). Specifically, under the electric field $E(\omega)$ of incident light, the second-order nonlinear light polarization $P(2\omega)$ with the frequency 2$\omega$ emerges via the third-rank SHG susceptibility coefficients $\chi _{abc}$: ${P^a}(2\omega ) = {\varepsilon _0}\chi _{abc}{E^b}(\omega ){E^c}(\omega )$, where $\varepsilon _0$ is the vacuum permittivity, the indices $a$, $b$, and $c$ indicate the Cartesian axes. According to the single particle length-gauge theory of nonlinear optical response\cite{A1995TH68200052,000072163300042}, the total ${\chi _{abc}}$ coefficients contain two components: ${\chi _{abc}} = \chi ^e_{abc} + \chi ^i_{abc}$, where $\chi ^e_{abc}$ represents the interband contribution, given by:
\begin{align}
	\chi ^e_{abc} = \frac{{{e^3}}}{{{\hbar ^2}\Omega }}\mathop \sum \limits_{nml} \frac{{r_{nm}^a\{ r_{ml}^br_{ln}^c\} }}{{({\omega _{ln}} - {\omega _{ml}})}} \times [\frac{{2{f_{nm}}}}{{{\omega _{mn}} - 2\omega }} + \frac{{{f_{ln}}}}{{{\omega _{ln}} - \omega }} + \frac{{{f_{ml}}}}{{{\omega _{ml}} - \omega }}]
\end{align}
and $\chi ^i_{abc}$ accounts for the mixed interband-intraband processes:
\begin{align}
	\chi ^i_{abc} = \frac{{i{e^3}}}{{2{\hbar ^2}\Omega }}\mathop \sum \limits_{nm} {f_{nm}}[\frac{2}{{{\omega _{mn}}({\omega _{mn}} - 2\omega )}}r_{nm}^a(r_{nm;c}^b + r_{mn;b}^c) + \frac{1}{{{\omega _{mn}}({\omega _{mn}} - 2\omega )}}(r_{nm;c}^br_{mn}^b + r_{nm;b}^ar_{mn}^c) \nonumber \\
	+ \frac{1}{{\omega _{mn}^2}}(\frac{1}{{{\omega _{mn}} - \omega }} - \frac{4}{{{\omega _{mn}} - 2\omega }})r_{nm}^a(r_{mn}^b\Delta _{mn}^c + r_{mn}^c\Delta _{mn}^b) - \frac{1}{{2{\omega _{mn}}({\omega _{mn}} - \omega )}}(r_{nm;a}^br_{mn}^c + r_{nm;a}^cr_{mn}^b)]
\end{align}
Except for $\omega$, all quantities appeared above depend on electron momentum $k$, while $k$ index is omitted for clarity. $r_{nm;a}^b$ is the generalized derivative of the coordinate operator in $k$ space, given by:
\begin{align}
	{(r_{nm}^b)_{;{k^a}}} = \frac{{r_{nm}^a\Delta _{mn}^b + r_{nm}^b\Delta _{mn}^a}}{{{\omega _{nm}}}} + \frac{i}{{{\omega _{nm}}}} \times \mathop \sum \limits_l ({\omega _{lm}}r_{nl}^ar_{lm}^b - {\omega _{nl}}r_{nl}^br_{lm}^a)
\end{align}
According to symmetry requirement impose by $R3$ space group($C_3$ point group), FE SnP$_2$S$_6$ (SnP$_2$Se$_6$) has six independent SHG susceptibility coefficients: $\chi_{xxx}$ = $-\chi_{xyy}$ = $-\chi_{yxy}$ = $-\chi_{yyx}$, $\chi_{yyy}$ = $-\chi_{xxy}$ = $-\chi_{xyx}$ = $-\chi_{yxx}$, $\chi_{xyz}$ = $\chi_{xzy}$, $\chi_{zxx}$, $\chi_{xxz}$ = $\chi_{xzx}$ and $\chi_{zzz}$ (see supporting information for detailed derivation). Moreover, as 2D FE SnP$_2$S$_6$ (SnP$_2$Se$_6$) trilayer has the exactly same crystallographic symmetry as 3D FE bulk. $\chi_{abc}$ coefficients for bulk FE phase will be simulated and analyzed. The experimentally measurable SHG intensity $I(2 \omega)$ strongly depends $\chi_{abc}$. In order to distinguish the polarization orientation of SnP$_2$S$_6$ (SnP$_2$Se$_6$) using SHG characterizations, the information about how symmetry adaptive $\chi_{abc}$ coefficients transform upon polarization switching of FE SnP$_2$S$_6$ (SnP$_2$Se$_6$) should be established.

\begin{figure}[!htb]
	\centering
	\includegraphics[width=1\linewidth]{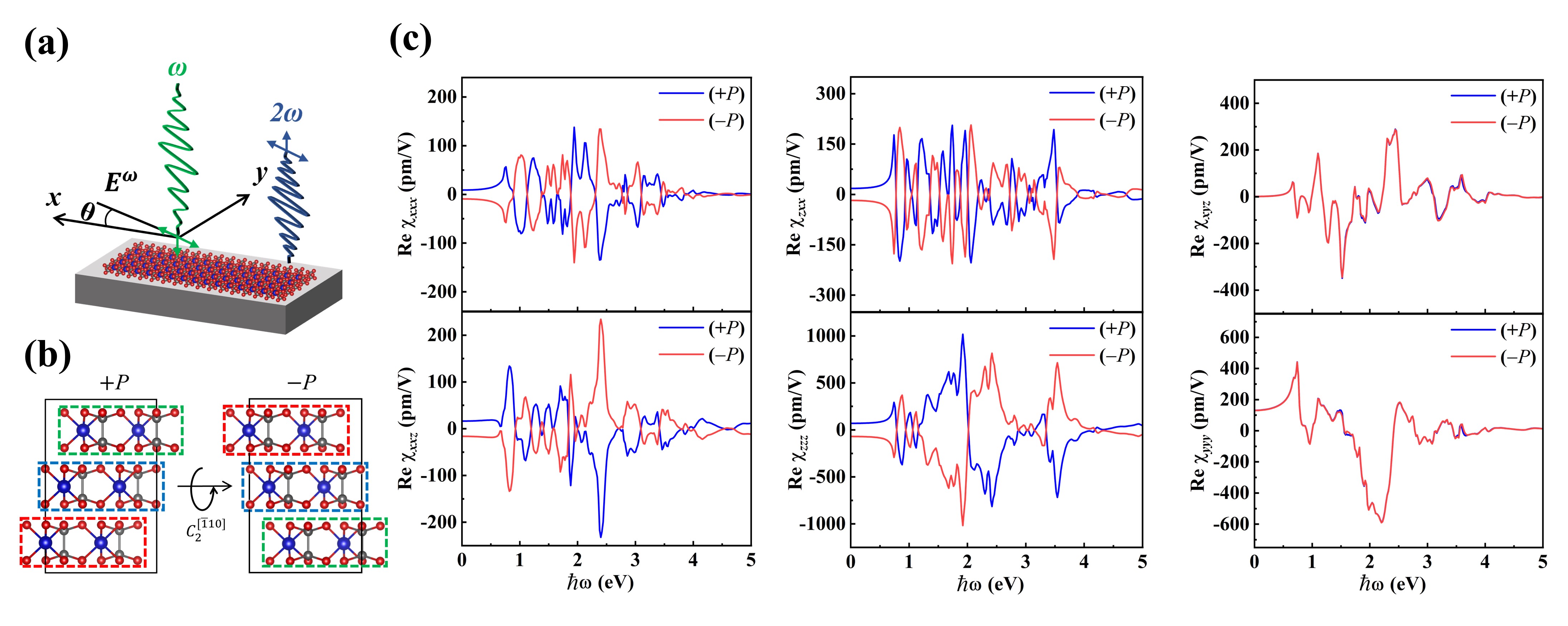}\vspace{-3pt}
	\caption{(a) Illustration of the SHG polarimetry geometry, where the linearly-polarized light with the frequency $\omega$ and polarization angle $\theta$ propagating along out-of-plane $z$ axis is incident on FE SnP$_2$S$_6$ among $xy$ Cartesian plane. (b) $\pm$$P$ states of SnP$_2$S$_6$ with opposite FE polarizations are interconnected by $C_2$ rotation operation applied along [$\overline1$10] direction. (c) The simulated real parts of six independent SHG susceptibility coefficients for FE SnP$_2$S$_6$ bulk. Blue and red lines represent $\chi(\omega)$ for $\pm$$P$ states, respectively. Except for $\chi _{xyz}$ and $\chi _{yyy}$, other coefficients reverse their signs upon polarizations switching.}
	\label{Figure 3}
\end{figure}

Figure 3b displays $\pm$$P$ states of FE-SnP$_2$S$_6$ bulk with opposite FE polarizations and reversed interlayer stacking sequences, which can be interconnected by two-fold rotation applied along [$\overline1$10] diagonal direction ($C_2^{[\overline110]}$). Under the Cartesian coordinates, such $C_2^{[\overline110]}$ operation can be represented by a second-rank tensor as:
\begin{align}
	D_{ij} = \left( {\begin{array}{*{20}{c}}
			{{D_{xx}}}&{{D_{xy}}}&{{D_{xz}}}\\
			{{D_{yx}}}&{{D_{yy}}}&{{D_{yz}}}\\
			{{D_{zx}}}&{{D_{zy}}}&{{D_{zz}}}
	\end{array}} \right) = \left( {\begin{array}{*{20}{c}}
			{\frac{1}{2}}&{ - \frac{{\sqrt 3 }}{2}}&0\\
			{ - \frac{{\sqrt 3 }}{2}}&{ - \frac{1}{2}}&0\\
			0&0&{ - 1}
	\end{array}} \right)
\end{align}
Polarization switching within FE SnP$_2$S$_6$ is equivalent to performing $C_2^{[\overline110]}$ operation, SHG susceptibility coefficients for $\pm$$P$ states therefore adopt the relation: $\chi _{ijk}^{ - {P}} = \sum\limits_{a,b,c} {{D_{ia}}{D_{jb}}{D_{kc}}\chi _{abc}^{ + {P}}} $. Using $\chi_{xxx}$ as an example, combining all nonzero $D_{ij}$ and $\chi_{abc}$ coefficients, we have:
\begin{equation}
	\begin{aligned}
		\chi_{xxx}^{-{P}} &= D_{xx} D_{xx} D_{xx} \chi_{xxx}^{+P} 
		+ D_{xy} D_{xy} D_{xy} \chi_{yyy}^{+P} 
		+ D_{xy} D_{xx} D_{xx} \chi_{yxx}^{+P} \\
		&\quad 
		+ D_{xx} D_{xy} D_{xy} \chi_{xyy}^{+P} 
		+ 2 D_{xy} D_{xy} D_{xx} \chi_{yyx}^{+P} 
		+ 2 D_{xx} D_{xx} D_{xy} \chi_{xxy}^{+P} \\
		&= \frac{1}{8} \chi_{xxx}^{+P} 
		- \frac{3\sqrt{3}}{8} \chi_{yyy}^{+P} 
		- \frac{\sqrt{3}}{8} \chi_{yxx}^{+P} 
		+ \frac{3}{8} \chi_{xyy}^{+P} 
		+ \frac{3}{4} \chi_{yyx}^{+P} 
		- \frac{\sqrt{3}}{4} \chi_{xxy}^{+P} \\
		&= \frac{1}{8} \chi_{xxx}^{+P} 
		- \frac{3\sqrt{3}}{8} \chi_{yyy}^{+P} 
		+ \frac{\sqrt{3}}{8} \chi_{yyy}^{+P} 
		- \frac{3}{8} \chi_{xxx}^{+P} 
		- \frac{3}{4} \chi_{xxx}^{+P} 
		+ \frac{\sqrt{3}}{4} \chi_{yyy}^{+P} \\
		&= -\chi_{xxx}^{+P}
	\end{aligned}
\end{equation}
Similarly, upon polarization switching, other five independent SHG susceptibility coefficients transform as: $\chi_{zxx}^{-{P}}=-\chi_{zxx}^{+{P}}$, $\chi_{xxz}^{-{P}}=-\chi_{xxz}^{+{P}}$, $\chi_{yyy}^{-{P}}=\chi_{yyy}^{+{P}}$, $\chi_{zzz}^{-{P}}=-\chi_{zzz}^{+{P}}$, $\chi_{xyz}^{-{P}}=\chi_{xyz}^{+{P}}$. According to symmetry analysis performed above, only $\chi_{xyz}$ and $\chi_{yyy}$ coefficients containing odd-number $y$ indices remain invariant under polarization switching of FE-SnP$_2$S$_6$ (SnP$_2$Se$_6$).

Beside the analytical derivation, computational simulations of all independent SHG susceptibility coefficients have also been performed based on PBE electronic structure (Figure S4). Shown in Figure 3c are the simulated real part of frequency dependent $\chi_{abc}(\omega)$ for $\pm$$P$ states of FE SnP$_2$S$_6$ bulk. Through the entire photo frequency range, $\pm$$P$ states have exactly same $\chi_{xyz}$ and $\chi_{yyy}$ coefficients, while other four coefficients reverse their signs after switching of polarization. Therefore, the transformation rules of $\chi_{abc}$ coefficients upon polarization switching derived from symmetry analysis are consistent with computational simulations. Specifically, the maximal SHG susceptibility magnitude up to 1000 pm/V appears in $\chi_{zzz}$ under the incident photon energy of 1.9 eV (0.5 eV above simulated $E_g$). Meanwhile, all six independent $\chi_{abc}$ coefficients exhibit the substantial magnitudes at photon energy below $E_g$, originating from the resonant contributions associated with interband transitions and non-resonant background responses arising from virtual transitions\cite{A1996UJ48500053}.

For better understanding of the macroscopic origin responsible for different transformation rules of SHG susceptibility coefficients (e.g. $\chi_{xxx}$ $vs$ $\chi_{yyy}$) upon polarization switching, tracing the $k$ dependent $\chi_{abc}$ coefficients of FE SnP$_2$S$_6$ are also performed. As illustrated by Figure 4a, $C_2^{[\overline110]}$ operation applied on SnP$_2$S$_6$ in real space is equivalent to performing $C_2$ rotation along $\Gamma-M$ path ($k_y$ axis) of reciprocal space. Recalling the $k$-dependent quantities appeared in Equation (1) - (3), only the vector terms containing $y$ index (e.g. $r_{nm}^y$ and $\Delta_{mn}^y$) are invariant under $C_2^{k_y}$, others vector terms with either $x$ or $z$ index will reverse their signs upon polarization switching (e. g. $r_{nm}^{x(z)} (-P) = - r_{nm}^{x(z)} (+P)$ and $\Delta_{mn}^{x(z)} (-P) = - \Delta_{mn}^{x(z)} (+P)$). Therefore, $\chi_{xxx}$ coefficient has its sign reversed upon polarization switching, whereas $\chi_{yyy}$ remains invariant. As shown in Figure 4b, the symmetry constraint imposed by $C_2^{k_y}$ operation are clearly visualized by the simulated $k$ dependent $\chi_{xxx}$ and $\chi_{yyy}$ coefficients for $\pm$$P$ states of SnP$_2$S$_6$ at $h\omega = 0.94$ eV (0.46 eV below E$_g$). Specifically, sign reversing of $\chi_{xxx}$ coefficient upon polarization switching ($\chi_{xxx}^{-P}$ = $C_2^{k_y}$$\chi_{xxx}^{+P}$ =  $-\chi_{xxx}^{+P}$) can be rationalized by red-to-blue color spot changing within the enlarged dashed areas connected by $C_2^{k_y}$ operation. While the sign (spot color) of $\chi_{yyy}$ remains unchanged under $C_2^{k_y}$ operation associated the polarization switching. 

\begin{figure}[!htb]
	\centering
	\includegraphics[width=1\linewidth]{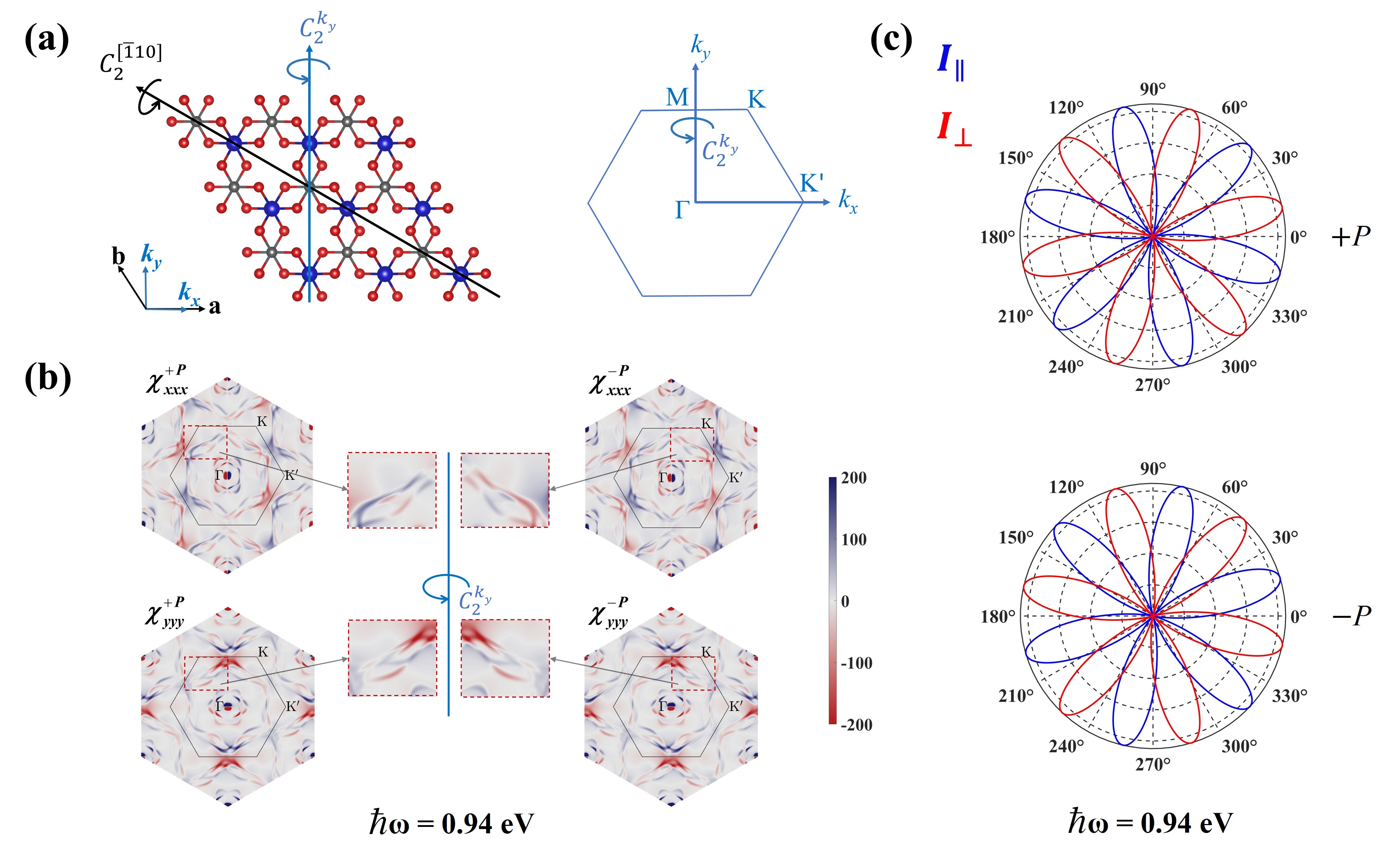}\vspace{-3pt}
	\caption{(a) Illustration of polarization switching in FE SnP$_2$S$_6$ via rotation operation performed in real space (black) and reciprocal space (blue). (b) The simulated real parts of $\chi_{xxx}$ and $\chi_{yyy}$ coefficients among the reciprocal space for $\pm$$P$ states of FE SnP$_{2}$S$_{6}$. The inset indicates how SHG susceptibility coefficients transform upon polarization switching and associated $C_2$ rotation operation. (c) The simulated SHG polar plots for $\pm$$P$ states of FE SnP$_2$S$_6$, where red and blue lines represent SHG responses perpendicular and  parallel to $E(\omega)$ of the incident linearly polarized light. At the characteristic frequency of $\hbar\omega = 0.94$ eV, the patterns of $I_\parallel$ and $I_\bot$ are nearly swapped between $\pm$$P$ states.}
	\label{Figure 4}
\end{figure}

The distinct transformation rules between $\chi_{xxx}$ and $\chi_{yyy}$ SHG coefficients of SnP$_2$S$_6$ have the direct consequences on SHG polar plot (angle resolved SHG intensity). Considering SHG polarimetry geometry illustrated in Figure 3a and the parallel back-scattering experimental setup\cite{000621232800002} commonly applied for 2D materials, $E(\omega)$ of incident linearly polarized light are dominated by the in-plane electric field $E_x = E_0 \cos\theta$ and $E_y = E_0 \sin\theta$, while $E(\omega)$ along out-of-plane $z$ axis is negligible ($E_z \cong 0$). In this way, the parallel ($\parallel$) and perpendicular ($\bot$) components of the second-order nonlinear light polarization with respect to $E(\omega)$ of incident light can be written as:
\begin{equation}
	\begin{aligned}
		P_\parallel &= P_x \cos\theta + P_y \sin\theta \\ P_\bot &= -P_x \sin\theta + P_y \cos\theta
	\end{aligned}
\end{equation}
where $\theta$ measures the angle between $E(\omega)$ and Catessian $x$ axis of SnP$_2$S$_6$. Considering light polarization components $P_x$ and $P_y$ can be expressed by: $P_x = \chi_{xxx} E_x^2 + \chi_{xyy} E_y^2 + (\chi_{xxy} + \chi_{xyx}) E_x E_y$ and $P_y = \chi_{yxx} E_x^2 + \chi_{yyy} E_y^2 + (\chi_{yxy} + \chi_{yyx}) E_x E_y$, the experimentally measurable SHG intensity along the parallel and perpendicular light polarization directions are given by:
\begin{equation}
	\begin{aligned}
		I_\parallel &\propto |P_\parallel|^2 =
		\left( 
		4\chi_{xxx} \cos^3\theta 
		- 3\chi_{xxx} \cos\theta 
		+ 4\chi_{yyy} \sin^3\theta 
		- 3\chi_{yyy} \sin\theta 
		\right)^2 \\
		I_\bot &\propto |P_\bot|^2 =
		\left( 
		-4\chi_{xxx} \sin^3\theta 
		+ 3\chi_{xxx} \sin\theta 
		+ 4\chi_{yyy} \cos^3\theta 
		- 3\chi_{yyy} \cos\theta 
		\right)^2
	\end{aligned}
\end{equation}\par

Figure 4c displays the simulated SHG polar plots for $+P$ state of SnP$_2$S$_6$. In agreement with experimental measurement\cite{000980689000021}, our simulated $I_\parallel$ and $I_\bot$ components adopt the six-fold polar patterns, where the characteristic angles corresponding to the maximal $I_\parallel$ and $I_\bot$ magnitudes differ by $30^\circ$. Since $\chi_{xxx}(\omega)$ and $\chi_{yyy}(\omega)$ coefficients contributing to SHG intensity of SnP$_2$S$_6$ undergo the distinct transformation rules upon polarization switching: $\chi_{xxx}(-P)$ = $-\chi_{xxx}(+P)$ , $\chi_{yyy}(-P)$ = $\chi_{yyy}(+P)$. Therefore, after switching of polarization, $\pm$$P$ states of SnP$_2$S$_6$ will exhibit SHG polar plots in same six-fold shape, but the SHG polar directions (indicated by the characteristic angles) change accordingly. Especially, under the condition where $|\chi_{xxx}(\omega)|$ = $|\chi_{yyy}(\omega)|$, $I_\parallel$ and $I_\bot$ for $\pm P$ states of SnP$_2$S$_6$ can be exactly swapped. In fact, above condition is almost ideally satisfied at frequency $\hbar\omega = 0.94$ eV. Moreover, FE-SnP$_2$Se$_6$ has the exactly same crystallographic symmetry and $C_2$ rotation induced polarization switching with SnP$_2$S$_6$, the nearly swapping of $I_\parallel$ and $I_\bot$ for FE bulk SnP$_2$Se$_6$ can be realized at $\hbar\omega = 1.92$ eV (see Figure S5 for details). Therefore, the remarkable dependence of SHG polar plots upon polarization switching in FE SnP$_2$S$_6$ and SnP$_2$Se$_6$, manifests as a direct experimental proof for distinguishing the orientation of sliding ferroelectricity via SHG characterizations.

\begin{figure}[!htb]
	\centering
	\includegraphics[width=1\linewidth]{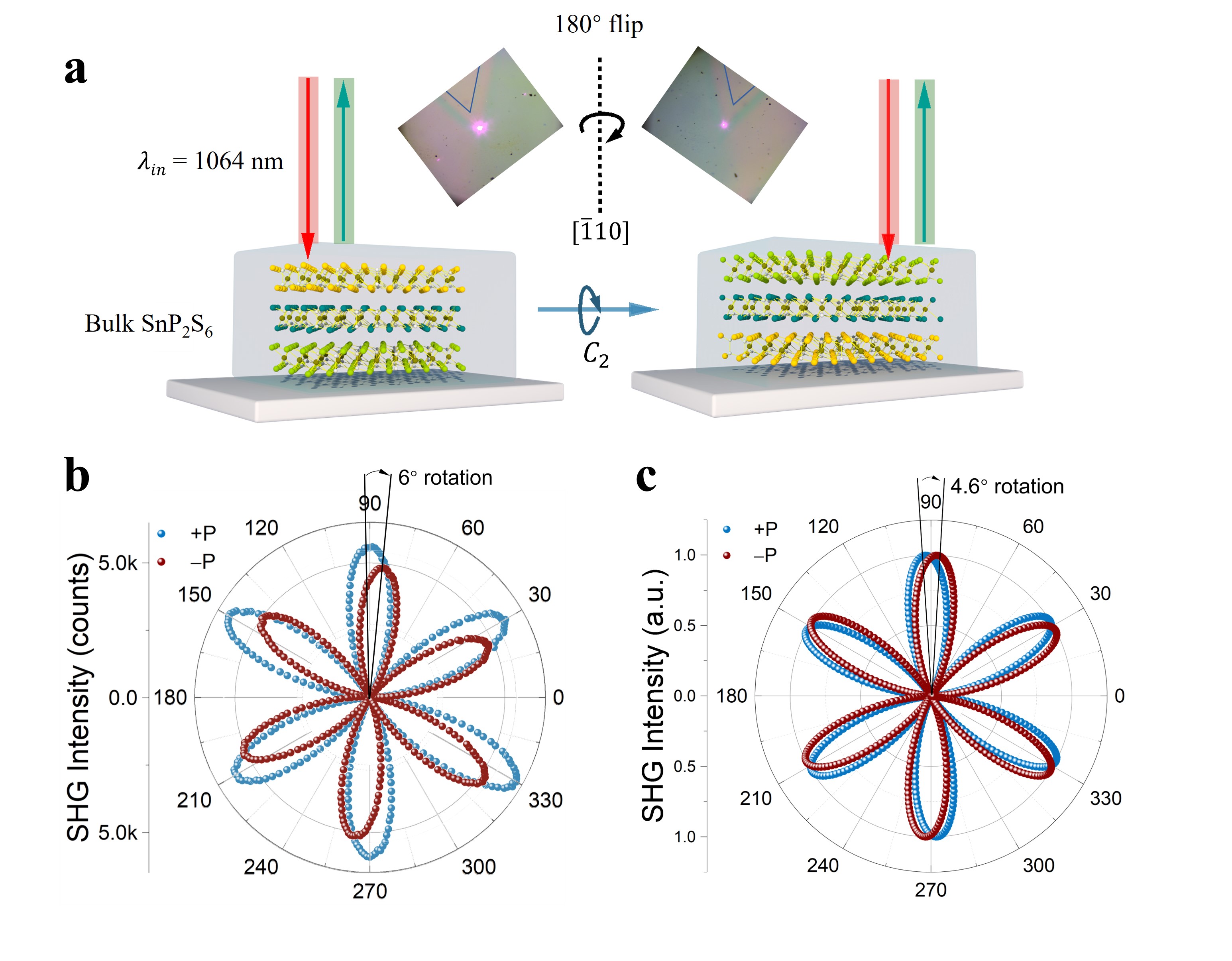}\vspace{-3pt}
	\caption{(a) Schematic diagram and optical morphologies for ferroelectricity switching of SnP$_2$S$_6$ through performing 180$^\circ$ flipping along the diagonal $[\overline110]$ direction. Pink dot represents the laser spot. (b) The measured polarization angle dependent SHG spectrum (containing only the parallel component $I_\parallel$) of SnP$_2$S$_6$ crystal incident by 1064 nm laser. After performing 180$^\circ$ flipping, about 6$^\circ$ rotation of SHG polar direction for SnP$_2$S$_6$ is detected. (c) The simulated SHG polar plots ($I_\parallel$ component only) for $\pm$$P$ states of FE SnP$_2$S$_6$ bulk under the photon energy of  $\hbar\omega = 0.74$ eV ($\approx$ 1/2 theoretical $E_g$), where the net rotation of SHG polar direction by 4.6$^\circ$ is predicted.}
	\label{Figure 5}
\end{figure}

The dependence of SHG polar directions on the ferroelectricity orientation of SnP$_2$S$_6$ is further validated by experimental measurement. To this end, SnP$_2$S$_6$ single crystal was prepared for characterizations. The measured XRD pattern, Raman and energy dispersive X-ray spectrometer (EDS) spectrum (see Figure S6 and Figure S7) indicate the high crystallinity of SnP$_2$S$_6$ crystal sample. As-synthesized SnP$_2$S$_6$ crystallizes  in a single-domain structural form, without any detectable FE domain-wall (see PFM images of SnP$_2$S$_6$ in Figure S8). Under the applied vertical electric field ($E_z$), switching out-of-plane ferroelectricity via in-plane sliding of individual monolayers requires the non-zero off-diagonal Born effective charge (BEC) tensors ($Z^*_{xz}$, $Z^*_{yz}$ $\neq$ 0), which interconnect the in-plane sliding driving forces with $E_z$ ($F_{x/y}$ = $Z^*_{xz/yz}$ $\times$ $E_z$)\cite{001491992000003,001501207500011}. However, for single-domain SnP$_2$S$_6$ crystal under hexagonal symmetry, its off-diagonal BEC tensors are exactly zero based on symmetry requirement ($Z^*_{xz/yz}$  $\equiv$ 0)\cite{001541333600002}. In other words, the out-of-plane polarization of SnP$_2$S$_6$ in single-domain structural form is not switchable under the vertical electric field, one have to reverse polarization orientation of SnP$_2$S$_6$ “manually”.

Based on our symmetry analysis, switching the ferroelectricity orientation of SnP$_2$S$_6$ is equivalent to performing $C_2^{[\overline110]}$ operation, we therefore “reverse” polarization of SnP$_2$S$_6$ through performing 180$^\circ$ flipping of SnP$_2$S$_6$ crystal along $[\overline110]$ crystallographic direction. As illustrated in Figure 5a, conducting the SHG polarimetry characterizations on the top surfaces of the (001)-oriented SnP$_2$S$_6$ crystal before and after 180$^\circ$ flipping, the dependence of SHG polar directions on the polarization orientation of SnP$_2$S$_6$ can be unambiguously identified. After careful correction of potential angular misalignment among $xy$ plane during the sample flipping, an obvious rotation of SHG polarization direction by approximately 6° is detected in the measured SHG polar plot pattern (Figure 5b) when SnP$_2$S$_6$ crystal is irradiated by 1064 nm laser. Performing SHG polar plot characterizations on the multiple SnP$_2$S$_6$ samples indicate a statistically averaged rotation angle of 5.84 $\pm$ 0.56 degree (Figure S9). Considering the photon energy of 1064 nm laser is approximately half of experiment band gap of SnP$_2$S$_6$ ($E_g$ = 2.3 eV), for consistency, SHG polar plots for $\pm$$P$ states of SnP$_2$S$_6$ bulk under the photo frequency of $\hbar\omega$ = 0.74 eV ($\approx$ 1/2 of theoretical $E_g$) have been simulated and compared directly to the experimental results. As shown in Figure 5c, a net rotation of SHG polar direction by 4.6$^\circ$ is predicted between $\pm$$P$ states of SnP$_2$S$_6$, in reasonable agreement with experimental measurement. Therefore, the remarkable  dependence of SHG polar directions on ferroelectricity orientation of SnP$_2$S$_6$ is validated by experiment.

Before conclusion, we will highlight the rareness of FE SnP$_2$S$_6$ (SnP$_2$Se$_6$) as sliding ferroelectrics for realization of polarization orientation dependent SHG responses. In order to distinguish the orientation of sliding ferroelectricity via SHG characterizations, the multiple independent SHG coefficients should contribute to the experimentally measurable $I_\parallel$ and $I_\bot$ components. In the meanwhile, these SHG coefficients should adopt the distinct transformation rules upon polarization switching. In other words, if all $\chi_{abc}$ coefficients involved in $I_\parallel$ and $I_\bot$ either remain unchanged, or simultaneously change all their signs after switching of polarization, then the SHG polar plots for $\pm$$P$ states will be identical, making it impossible to distinguish polarization orientation. Following the similar symmetry analysis we performed on FE-SnP$_2$S$_6$, we also derive how $I_\parallel$ and $I_\bot$ transform between $\pm$$P$ states for typical sliding ferroelectrics synthesized by experiment so far. As tabulated in TABLE I, for sliding ferroelectrics with $C_{2v}$ point group, all five independent $\chi_{abc}$ coefficients have their signs reversed when polarizations are switched between $\pm$$P$ states. As a result, SHG polar plots for $\pm$$P$ states remain unchanged under the proper SHG polarimetry geometry (i.e. $E_y = 0$, $E_{x,z} \neq$ 0). Meanwhile, for sliding ferroelectrics such as 3R-MoS$_2$, 3R-BN and $\gamma$-InSe in $C_{3v}$ point group, their  $I_\parallel$ and $I_\bot$ SHG components contain only one independent SHG coefficient $\chi_{yyy}$ under the parallel back-scattering SHG polarimetry. No matter how $\chi_{yyy}$ transforms upon polarization switching, SHG polar plots for $\pm$$P$ states are exactly same. Therefore, for most currently known sliding ferroelectrics, their SHG polar plots remain unchanged when polarizations are switched between $\pm$$P$ states, making it unable to distinguish the orientation of sliding ferroelectricity via SHG characterizations. Based on the symmetry analysis, only FE SnP$_2$S$_6$ (SnP$_2$Se$_6$) and other potential sliding ferroelectrics in $C_3$ point group can display the distinguishable SHG responses dependent on their ferroelectricity orientations.

\begin{table}[H]
	\centering
	\caption{For the known sliding ferroelectrics with either $C_{2v}$ or $C_{3v}$ polar point group, two-fold rotation along the interlayer sliding directions required to switch FE polarizations are provided, transformation rules for independent SHG coefficients and SHG intensity ($I_\parallel$ and $I_\bot$) between $\pm P$ states are also derived. For sliding ferroelectrics with $C_{2v}$ point group, their SHG intensities are exactly zero under the parallel back-scattering setup, therefore the alternative SHG polarimetry geometry is provided to obtain the nonzero $I_\parallel$ and $I_\bot$.}
	\scalebox{0.75}{
		\begin{tabular}{ccc@{\hspace{15pt}}ccc}
			\hline \hline
			\makecell{Prototypical sliding \\ ferroelectrics} & \makecell{Symmetry \\ operation} & \makecell{Independent SHG \\ coefficients ($+P$)} & \makecell{Independent SHG \\ coefficients ($-P$)} & \makecell{SHG \\ intensity} & \makecell{Orientation \\ distinguishable$?$}  \\
			\hline
			\makecell{1T'-WTe$_2$ \\ Td-MoTe$_2$ in $C_{2v}$} 
			& \makecell{$C_2^{[010]}$ \\ 
				$\begin{pmatrix}
					-1 & 0 & 0 \\
					0 & 1 & 0 \\
					0 & 0 & -1
				\end{pmatrix}$ \\  \\} 
			& \makecell{$\chi_{xzx}$($\chi_{xxz}$) \\ $\chi_{yyz}$($\chi_{yzy}$) \\ $\chi_{zxx}$ \\ $\chi_{zyy}$ \\ $\chi_{zzz}$ \\  \\} 
			& \makecell{$-\chi_{xzx}$($-\chi_{xxz}$) \\ $-\chi_{yyz}$($-\chi_{yzy}$) \\ $-\chi_{zxx}$ \\ $-\chi_{zyy}$ \\ $-\chi_{zzz}$ \\  \\} 
			& \makecell{\\ \textbf{$E_x = 0$, $E_{y,z} \neq$ 0}\\ [0.5em] 
				$\begin{aligned}
					I_\parallel &\propto \bigl[\chi_{zzz} \sin^3\theta \\
					&\quad + (\chi_{zyy} + 2\chi_{yyz}) \cos^2\theta \sin\theta \bigr]^2 \\
					I_\perp &\propto \bigl[\chi_{zyy} \cos^3\theta \\
					&\quad + (\chi_{zzz} - 2\chi_{yyz}) \sin^2\theta \cos\theta \bigr]^2
				\end{aligned}$} 
			& \makecell{\\ No} \\
			\hline
			\makecell{ \\ 3R-MoS$_2$ \\ 3R-BN \\ $\gamma$-InSe in $C_{3v}$} 
			& \makecell{ \\ $C_2^{[\overline110]}$ \\ 
				$\begin{pmatrix}
					0 & -1 & 0 \\
					-1 & 0 & 0 \\
					0 & 0 & -1
				\end{pmatrix}$} 
			& \makecell{ \\ $\chi_{xzx}$($\chi_{xxz}$) \\ $\chi_{yyy}$ \\ $\chi_{zxx}$ \\ $\chi_{zzz}$} 
			& \makecell{ \\ $-\chi_{xzx}$($-\chi_{xxz}$) \\ $\chi_{yyy}$ \\ $-\chi_{zxx}$ \\ $-\chi_{zzz}$} 
			& \makecell{ \\ \textbf{$E_z = 0$, $E_{x,y} \neq$ 0} \\[0.5em]
				$\begin{aligned}
					I_\parallel &\propto \bigl[\chi_{yyy} \sin^3\theta \\
					&\quad - 3\chi_{yyy} \cos^2\theta \sin\theta \bigr]^2 \\
					I_\perp &\propto \bigl[-\chi_{yyy} \cos^3\theta \\
					&\quad + 3\chi_{yyy} \sin^2\theta \cos\theta \bigr]^2
				\end{aligned}$ \\  \\} 
			& \makecell{\\ \\ No} \\
			\hline
			\makecell{ SnP$_2$S$_6$ \\ SnP$_2$Se$_6$ in $C_{3}$} 
			& \makecell{$C_2^{[\overline110]}$ \\ 
				$\begin{pmatrix}
					0 & -1 & 0 \\
					-1 & 0 & 0 \\
					0 & 0 & -1
				\end{pmatrix}$} 
			& \makecell{ $\chi_{xxx}$ \\ $\chi_{xxz}$($\chi_{xzx}$) \\ $\chi_{xyz}$($\chi_{xzy}$) \\ $\chi_{yyy}$ \\ $\chi_{zxx}$ \\ $\chi_{zzz}$ \\  \\} 
			& \makecell{ $-\chi_{xxx}$ \\ $-\chi_{xxz}$($-\chi_{xzx}$) \\ $\chi_{xyz}$($\chi_{xzy}$) \\ $\chi_{yyy}$ \\ $-\chi_{zxx}$ \\ $-\chi_{zzz}$ \\  \\} 
			& \makecell{\textbf{$E_z = 0$, $E_{x,y} \neq$ 0} \\[0.5em] 
				$\begin{aligned}
					I_\parallel &\propto \bigl[4\chi_{xxx} \cos^3\theta - 3\chi_{xxx} \cos\theta \\
					&\quad + 4\chi_{yyy} \sin^3\theta - 3\chi_{yyy} \sin\theta \bigr]^2 \\
					I_\perp &\propto \bigl[-4\chi_{xxx} \sin^3\theta + 3\chi_{xxx} \sin\theta \\
					&\quad + 4\chi_{yyy} \cos^3\theta - 3\chi_{yyy} \cos\theta  \bigr]^2
				\end{aligned}$} 
			& \makecell{\\ \\ Yes} \\
			\hline\hline
		\end{tabular}
	}
\end{table}

To summarize, after performing the computational simulations and experimental SHG characterizations, we demonstrate SnP$_2$S$_6$, 2D vertically polarized materials with strong SHG responses, as a new group of sliding ferroelectrics whose ferroelectricity orientation is distinguishable via SHG characterizations. Specifically, our first-principles calculations clearly indicate that out-of-plane polarizations of SnP$_2$S$_6$ and SnP$_2$Se$_6$ are direct consequence of rhombohedral $ABC$ stacking of individual monolayers without involving cation polar displacement, and switching of their FE polarizations can be realized via relative sliding of adjacent monolayers. Moreover, the only nontrivial symmetry operation (three-fold rotation) within FE SnP$_2$S$_6$ (SnP$_2$Se$_6$) gives arise to the multiple in-plane SHG susceptibility coefficients $\chi_{xxx}$ and $\chi_{yyy}$ contributing to the experimentally detectable SHG responses. Especially, incident by the linearly polarized light with a characteristic photo frequency, the ideal intersection between $\chi_{xxx}$ and $\chi_{yyy}$ coefficients with opposite signs and almost identical magnitudes, leads to a maximal $30^\circ$ rotation of SHG polar direction upon polarization switching of FE SnP$_2$S$_6$ (SnP$_2$Se$_6$). The dependence of SHG polar directions on the polarization orientation of FE SnP$_2$S$_6$ is further validated by experimental characterizations, refining the conventional understanding that SHG characterization is unable to distinguish ferroelectricity orientation. Our work not only demonstrates a new paradigm for effective and in-situ detecting the orientation of sliding ferroelectricity, but also establishes the controllable second-order optical nonlinearity for 2D sliding ferroelectrics with unique crystallographic symmetry.
	
G.Y.G. acknowledges the funding support from the National Natural Science Foundation of China (Grant No. 11574244). H.W. acknowledges the support from the NSFC under Grants Nos. 12304049 and 12474240, as well as the support provided by the Zhe-jiang Provincial Natural Science Foundation of China under grant number LDT23F04014F01. L.Y. acknowledges the support by the National Natural Science Foundation of China (No. 12474089). Hefei Advanced Computing Center is acknowledged for computational support.

\bibliography{reference}

\end{document}


\title{Distinguish the Orientation of Sliding Ferroelectricity by Second-Harmonic Generation}

\author{Fengfeng Ye}
\altaffiliation{These authors contributed equally to this work}
\affiliation{Frontier Institute of Science and Technology, State Key Laboratory of Electrical Insulation and Power Equipment, Xi’an Jiaotong University, Xi’an 710049, People's Republic of China}

\author{Qiankun Li}
\altaffiliation{These authors contributed equally to this work}
\affiliation{School of Physical Science and Technology, Jiangsu Key Laboratory of Frontier Material Physics and Devices, Soochow University, Suzhou, 215006, China}

\author{Zhuocheng Lu}
\altaffiliation{These authors contributed equally to this work}
\affiliation{Center for Quantum Matter, School of Physics, Zhejiang University, Zhejiang 310058, China}

\author{Xinfeng Chen}
\affiliation{Frontier Institute of Science and Technology, State Key Laboratory of Electrical Insulation and Power Equipment, Xi’an Jiaotong University, Xi’an 710049, People's Republic of China}

\author{Yang Li}
\affiliation{School of Materials Science and Engineering, Harbin Institute of Technology, Harbin 150001, China}

\author{Hua Wang}
\email{daodaohw@zju.edu.cn}
\affiliation{Center for Quantum Matter, School of Physics, Zhejiang University, Zhejiang 310058, China}

\author{Lu You}
\email{lyou@suda.edu.cn}
\affiliation{School of Physical Science and Technology, Jiangsu Key Laboratory of Frontier Material Physics and Devices, Soochow University, Suzhou, 215006, China}

\author{Gaoyang Gou}
\email{gougaoyang@mail.xjtu.edu.cn}
\affiliation{Frontier Institute of Science and Technology, State Key Laboratory of Electrical Insulation and Power Equipment, Xi’an Jiaotong University, Xi’an 710049, People's Republic of China}

\date{\today}

\maketitle
\clearpage

SnP$_2$X$_6$ (X = S and Se) adopts $C_3$ point group, which belongs to one of 10 polar point groups exhibiting SHG effect. The second-order nonlinear light polarization within SHG effect can be described as ${P^a}(2\omega ) = {\varepsilon _0}\chi _{abc}^{(2)}{E^b}(\omega ){E^c}(\omega )$. Both $P$ and $E$ are polar vectors, their transformation rules under $C_3$ point group adopt the relation: $\Gamma_P=\Gamma_E=A+E$. For linear susceptibility, we have $\Gamma_P\otimes\Gamma_E = 3A + 3E = (2A + 2E)^s + (A + E)^a$, where $s$ and $a$ represent the symmetric and antisymmetric parts, respectively. Without considering permutation symmetry, three nonzero coefficients arise from the three irreducible representations ($A$), so the linear susceptibility $\chi _{abc}^{(1)}$ has three independent coefficients. When permutation symmetry is considered, the symmetric part with two irreducible representations ($A$) results in two independent coefficients for the linear susceptibility $\chi _{abc}^{(1)}$.\par

For second-order nonlinear optical susceptibilities such as SHG, the corresponding representation is given by the direct product ${\Gamma _P} \otimes {\Gamma _E} \otimes {\Gamma _E}$. However, due to the permutation symmetry of the electric fields $E^b (\omega)$ and $E^c (\omega)$, and the direct product ${\Gamma _P} \otimes {\Gamma _E} \otimes {\Gamma _E}$ contains the representations of all tensor components, owing to the inherent permutation symmetry required by $E^b (\omega)$ and $E^c (\omega)$, the antisymmetric part is then discarded. Calculating the direct product of the symmetric part ${\Gamma _s} = 2A + 2E$ with the second electric field ${\Gamma _E}$, we obtain ${\Gamma _P} \otimes {\Gamma _E} \otimes {\Gamma _E} = {\Gamma _s} \otimes {\Gamma _E} = 6A + 6E$. Six irreducible representations ($A$) indicate that, without considering Kleinman symmetry, SnP$_2$X$_6$ (X = S and Se) under $C_3$ point group characterize six independent SHG susceptibility coefficients.\par

The six independent coefficients of SnP$_2$X$_6$ can be further determined by tensor analysis. Under the symmetry constraints of $C_{3}$ point group, each matrix element of SHG susceptibility undergoes a transformation imposed by the corresponding symmetry operation, whose numerical values should remain invariant after the transformation. Under Cartesian coordinates, the three fold rotation operation along out-of-plane axis within $C_{3}$ point group can be described by the following 3 $\times$ 3 matrix as:
\begin{align}
	{A_{ij}} = \left( {\begin{array}{*{20}{c}}
			{{A_{xx}}}&{{A_{xy}}}&{{A_{xz}}}\\
			{{A_{yx}}}&{{A_{yy}}}&{{A_{yz}}}\\
			{{A_{zx}}}&{{A_{zy}}}&{{A_{zz}}}
	\end{array}} \right) = \left( {\begin{array}{*{20}{c}}
			{ - \frac{1}{2}}&{\frac{{\sqrt 3 }}{2}}&0\\
			{ - \frac{{\sqrt 3 }}{2}}&{ - \frac{1}{2}}&0\\
			0&0&1
	\end{array}} \right)
\end{align}
After performing three fold rotation operation, SHG susceptibility coefficients $\chi_{zxx}$, $\chi_{zyy}$ and $\chi_{zyx}$ adopt the following transformation relations(in Einstein summation notation):
\begin{align}
	{\chi _{zxx}} & = {A_{za}}{A_{xb}}{A_{xc}}{\chi _{abc}} = {A_{zz}}{A_{xx}}{A_{xx}}{\chi _{zxx}} + {A_{zz}}{A_{xy}}{A_{xy}}{\chi _{zyy}} + {A_{zz}}{A_{xy}}{A_{xx}}{\chi _{zyx}} + {A_{zz}}{A_{xx}}{A_{xy}}{\chi _{zxy}} \nonumber \\
	& = \frac{1}{4}{\chi _{zxx}} + \frac{3}{4}{\chi _{zyy}} - \frac{{\sqrt 3 }}{2}{\chi _{zyx}}
\end{align}
\begin{align}
	{\chi _{zyy}} & = {A_{za}}{A_{yb}}{A_{yc}}{\chi _{abc}} = {A_{zz}}{A_{yx}}{A_{yx}}{\chi _{zxx}} + {A_{zz}}{A_{yy}}{A_{yy}}{\chi _{zyy}} + {A_{zz}}{A_{yy}}{A_{yx}}{\chi _{zyx}} + {A_{zz}}{A_{yx}}{A_{yy}}{\chi _{zxy}} \nonumber \\
	& = \frac{3}{4}{\chi _{zxx}} + \frac{1}{4}{\chi _{zyy}} + \frac{{\sqrt 3 }}{2}{\chi _{zyx}}
\end{align}
\begin{align}
	{\chi _{zyx}} & = {A_{za}}{A_{yb}}{A_{xc}}{\chi _{abc}} = {A_{zz}}{A_{yx}}{A_{xx}}{\chi _{zxx}} + {A_{zz}}{A_{yy}}{A_{xy}}{\chi _{zyy}} + {A_{zz}}{A_{yy}}{A_{xx}}{\chi _{zyx}} + {A_{zz}}{A_{yx}}{A_{xy}}{\chi _{zxy}} \nonumber \\
	& = \frac{{\sqrt 3 }}{4}{\chi _{zxx}} - \frac{{\sqrt 3 }}{4}{\chi _{zyy}} - \frac{1}{2}{\chi _{zyx}}
\end{align}
As above SHG susceptibility coefficients remain invariant under three-fold rotation operations, thus we have:
\begin{align}
	\frac{1}{4}{\chi_{zxx}} + \frac{3}{4}{\chi_{zyy}} - \frac{{\sqrt3}}{2}{\chi_{zyx}} = {\chi_{zxx}}
\end{align}
\begin{align}	
	\frac{3}{4}{\chi_{zxx}} + \frac{1}{4}{\chi_{zyy}} + \frac{{\sqrt3}}{2}{\chi_{zyx}} = {\chi_{zyy}}
\end{align}
\begin{align}	
	\frac{{\sqrt3}}{4}{\chi_{zxx}} - \frac{{\sqrt3}}{4}{\chi_{zyy}} - \frac{1}{2}{\chi_{zyx}} = {\chi_{zyx}}
\end{align}
By solving the ternary equations above, we can obtain ${\chi _{zyy}} = {\chi _{zxx}}$, ${\chi _{zxy}} = {\chi _{zyx}} = 0$. After performing similar symmetry operation analysis for the remaining $\chi_{abc}$ coefficients, the third-rank tensor for SHG susceptibility of FE SnP$_2$S$_6$ (SnP$_2$Se$_6$) can be described as:
\[\left( {\begin{array}{*{20}{c}}
		{{\chi _{xxx}}}&{ - {\chi _{xxx}}}&0&{{\chi _{xyz}}}&{{\chi _{xzy}}}&{{\chi _{xzx}}}&{{\chi _{xxz}}}&{ - {\chi _{yyy}}}&{ - {\chi _{yyy}}}\\
		{ - {\chi _{yyy}}}&{{\chi _{yyy}}}&0&{{\chi _{xxz}}}&{{\chi _{xzx}}}&{ - {\chi _{xzy}}}&{ - {\chi _{xyz}}}&{ - {\chi _{xxx}}}&{ - {\chi _{xxx}}}\\
		{{\chi _{zxx}}}&{{\chi _{zxx}}}&{{\chi _{zzz}}}&0&0&0&0&0&0
\end{array}} \right)\]
Clearly, FE-SnP$_2$S$_6$ (SnP$_2$Se$_6$) in $C_3$ point group contains six independent SHG susceptibility coefficients: $\chi _{xxx}$, ${\chi _{xyz}}({\chi _{xzy}})$, ${\chi _{zxx}}$, ${\chi _{xxz}}({\chi _{xzx}})$, ${\chi _{yyy}}$, ${\chi _{zzz}}$.\par

\begin{table}[H]
	\centering
	\caption{Our simulated crystallographic parameters for FE and PE phases of bulk SnP$_{2}$X$_{6}$(X=S, Se) in hexagonal setting. For the FE-SnP$_{2}$S$_{6}$, comparison between our calculated (Cal.) and experimental (Exp) crystallographic parameters are also presented.}
	\small
	\begin{tabular}{ccccccc|cccc}
		\hline \hline
		\multicolumn{7}{c|}{FE-SnP$_{2}$S$_{6}$ (R3)} & \multicolumn{4}{c}{FE-SnP$_{2}$Se$_{6}$ (R3)} \\
		\multicolumn{7}{c|}{Cal: $a$=$b$=5.986 Å, $c$=19.177 Å} & \multicolumn{4}{c}{Cal: $a$=$b$=6.320 Å, $c$=19.548 Å} \\
		\multicolumn{7}{c|}{Exp: $a$=$b$=5.999 Å, $c$=19.424 Å} & & & & \\
		\hline
		& \multicolumn{3}{c}{Cal} & \multicolumn{3}{c|}{Exp} & & \multicolumn{3}{c}{Cal} \\
		\hline
		Atom & $x$ & $y$ & $z$ & $x$ & $y$ & $z$ & Atom & $x$ & $y$ & $z$ \\
		\hline
		S (9b) & 0.0030 & 0.3287 & 0.1424 & 0.0095 & 0.3306 & 0.1407 & Se (9b) & 0.0045 & 0.3364 & 0.8888 \\
		\hline
		S (9b) & 0.3404 & 0.3340 & 0.3032 & 0.3378 & 0.3274 & 0.3045 & Se (9b) & 0.3378 & 0.3438 & 0.0449 \\
		\hline
		P (3a) & 0.0000 & 0.0000 & $-0.0016$ & 0.0000 & 0.0000 & 0.0000 & P (3a) & 0.0000 & 0.0000 & 0.7435 \\
		\hline
		P (3a) & 0.0000 & 0.0000 & 0.1145 & 0.0000 & 0.0000 & 0.1138 & P (3a) & 0.0000 & 0.0000 & 0.8588 \\
		\hline
		Sn (3a) & 0.0000 & 0.0000 & 0.3891 & 0.0000 & 0.0000 & 0.3894 & Sn (3a) & 0.0000 & 0.0000 & 0.1322 \\
		\hline \hline
		
		\multicolumn{7}{c|}{PE-SnP$_{2}$S$_{6}$ (P312)} & \multicolumn{4}{c}{PE-SnP$_{2}$Se$_{6}$ (P312)} \\
		\multicolumn{7}{c|}{Cal: $a$=$b$=5.983 Å, $c$=19.288 Å} & \multicolumn{4}{c}{Cal: $a$=$b$=6.310 Å, $c$=19.676 Å} \\
		\hline
		Atom & \multicolumn{6}{c|}{Cal} & Atom & \multicolumn{3}{c}{Cal} \\
		\hline
		S (6l) & \multicolumn{2}{c}{$-0.0007$} & \multicolumn{2}{c}{0.3269} & \multicolumn{2}{c|}{0.1384} & Se (6l) & 0.9886 & 0.3308 & 0.9194 \\
		\hline
		S (6l) & \multicolumn{2}{c}{0.0011} & \multicolumn{2}{c}{0.3273} & \multicolumn{2}{c|}{0.4749} & Se (6l) & $-0.0082$ & 0.3309 & 0.2559 \\
		\hline
		S (6l) & \multicolumn{2}{c}{0.3351} & \multicolumn{2}{c}{0.9947} & \multicolumn{2}{c|}{0.8068} & Se (6l) & 0.3263 & 0.9991 & 0.5880 \\
		\hline
		P (2g) & \multicolumn{2}{c}{0.0000} & \multicolumn{2}{c}{0.0000} & \multicolumn{2}{c|}{0.9953} & P (2g) & 0.0000 & 0.0000 & 0.7756 \\
		\hline
		P (2g) & \multicolumn{2}{c}{0.0000} & \multicolumn{2}{c}{0.0000} & \multicolumn{2}{c|}{0.3314} & P (2g) & 0.0000 & 0.0000 & 0.8898 \\
		\hline
		P (2h) & \multicolumn{2}{c}{0.3333} & \multicolumn{2}{c}{0.6667} & \multicolumn{2}{c|}{0.6631} & P (2h) & 0.3333 & 0.6667 & 0.4425 \\
		\hline
		Sn (2h) & \multicolumn{2}{c}{0.3333} & \multicolumn{2}{c}{0.6667} & \multicolumn{2}{c|}{0.3891} & Sn (2h) & 0.3333 & 0.6667 & 0.8318 \\
		\hline
		Sn (1f) & \multicolumn{2}{c}{0.6667} & \multicolumn{2}{c}{0.3333} & \multicolumn{2}{c|}{0.7210} & Sn (1f) & 0.6667 & 0.3333 & 0.5000 \\
		\hline \hline
	\end{tabular}
\end{table}

\begin{table}[H]
	\centering
	\caption{The overall FE polarization magnitudes for FE SnP$_2$X$_6$ (X = S and Se) bulk calculated using the Berry phase method, and estimated displacive polarization magnitudes based on the Born effecitve charge and Sn cation polar displacement.}
	\begin{tabular}{ccccc}
		\hline \hline
		\multicolumn{5}{c}{ SnP$_2$S$_6$ } \\
		\hline
		\makecell{Octahedron} & \makecell{Born effective charge (e)} & \makecell{Cation polar displacement (Å)} & \makecell{Direction} & \makecell{Polarization ($\mu$C/cm$^{2}$)} \\
		\hline
		[SnS$_{6}$] & Sn: 2.67 & 0.0083 & $\downarrow$ & $-0.18$ \\
		\hline
		\multicolumn{5}{c}{Estimated displacive polarization: $-0.18$ $\mu$C/cm$^{2}$ } \\
		\hline
		\multicolumn{5}{c}{Berry phase predicted polarization (bulk): 0.38 $\mu$C/cm$^{2}$ } \\
		\hline\hline
		\multicolumn{5}{c}{ SnP$_2$Se$_6$ } \\
		\hline
		\makecell{Octahedron} & \makecell{Born effective charge (e)} & \makecell{Cation polar displacement (Å)} & \makecell{Direction} & \makecell{Polarization ($\mu$C/cm$^{2}$)} \\
		\hline
		[SnSe$_{6}$] & Sn: 3.21 & 0.0247 & $\downarrow$ & $-0.57$ \\
		\hline
		\multicolumn{5}{c}{Estimated displacive polarization: $-0.57$ $\mu$C/cm$^{2}$ } \\
		\hline
		\multicolumn{5}{c}{Berry phase predicted polarization (bulk): 0.53 $\mu$C/cm$^{2}$ } \\
		\hline\hline
	\end{tabular}
\end{table}

\begin{figure}[!htb]
	\centering
	\includegraphics[width=1\linewidth]{FIG/1.jpg}\vspace{-3pt}
	\caption{ (a) The alternative polarization switching pathways (model II and III) for SnP$_2$S$_6$ trilayer, which involve the simultaneous sliding two of three monolayers. Change of (b) system energy and (c) polarization magnitudes along the above two polarization switching pathways. }
	\label{Figure S1}
\end{figure}

\begin{figure}[!htb]
	\centering
	\includegraphics[width=1\linewidth]{FIG/2.jpg}\vspace{-3pt}
	\caption{ The simulated potential energy surface (PES) associated with the relative sliding between two adjacent SnP$_2$Se$_6$ monolayers. }
	\label{Figure S2}
\end{figure}

\begin{figure}[!htb]
	\centering
	\includegraphics[width=1\linewidth]{FIG/3.jpg}\vspace{-3pt}
	\caption{ SnP$_{2}$Se$_{6}$ is a structural analogue of SnP$_{2}$S$_{6}$, displaying the similar interlayer sliding and polarization switching characters. Due to the weaker electronegativity of Se compared to S, and stronger interlayer electronic coupling of SnP$_2$Se$_6$ over SnP$_2$S$_6$, SnP$_2$Se$_6$ characterizes the stronger polarization magnitude and larger polarization switching barrier than SnP$_2$S$_6$. (a) Energy and (b) polarization magnitudes along the three polarization switching pathways predicted for SnP$_2$Se$_6$ trilayer. }
	\label{Figure S3}
\end{figure}

\begin{figure}[!htb]
	\centering
	\includegraphics[width=0.8\linewidth]{FIG/4.jpg}\vspace{-3pt}
	\caption{ (a) The band structures of FE SnP$_2$S$_6$ bulk simulated by PBE, predicting an energy band gap of 1.4 eV. (b) The band structure predicted by the tight-binding model is consistent with the PBE results. }
	\label{Figure S4}
\end{figure}

\begin{figure}[!htb]
	\centering
	\includegraphics[width=1\linewidth]{FIG/5.jpg}\vspace{-3pt}
	\caption{ (a) The simulated SHG susceptibility coefficients $\chi_{xxx}$ and $\chi_{yyy}$ for $\pm$$P$ states of FE-SnP$_2$Se$_6$ bulk that contribute its SHG polar plot under the parallel back-scattering setup. (b) The simulated  SHG polar plots for $\pm$$P$ states of SnP$_2$Se$_6$ under the incident linearly polarized monochromatic light with frequency $\hbar\omega$ = 1.92 eV. }
	\label{Figure S5}
\end{figure}

\begin{figure}[!htb]
	\centering
	\includegraphics[width=1\linewidth]{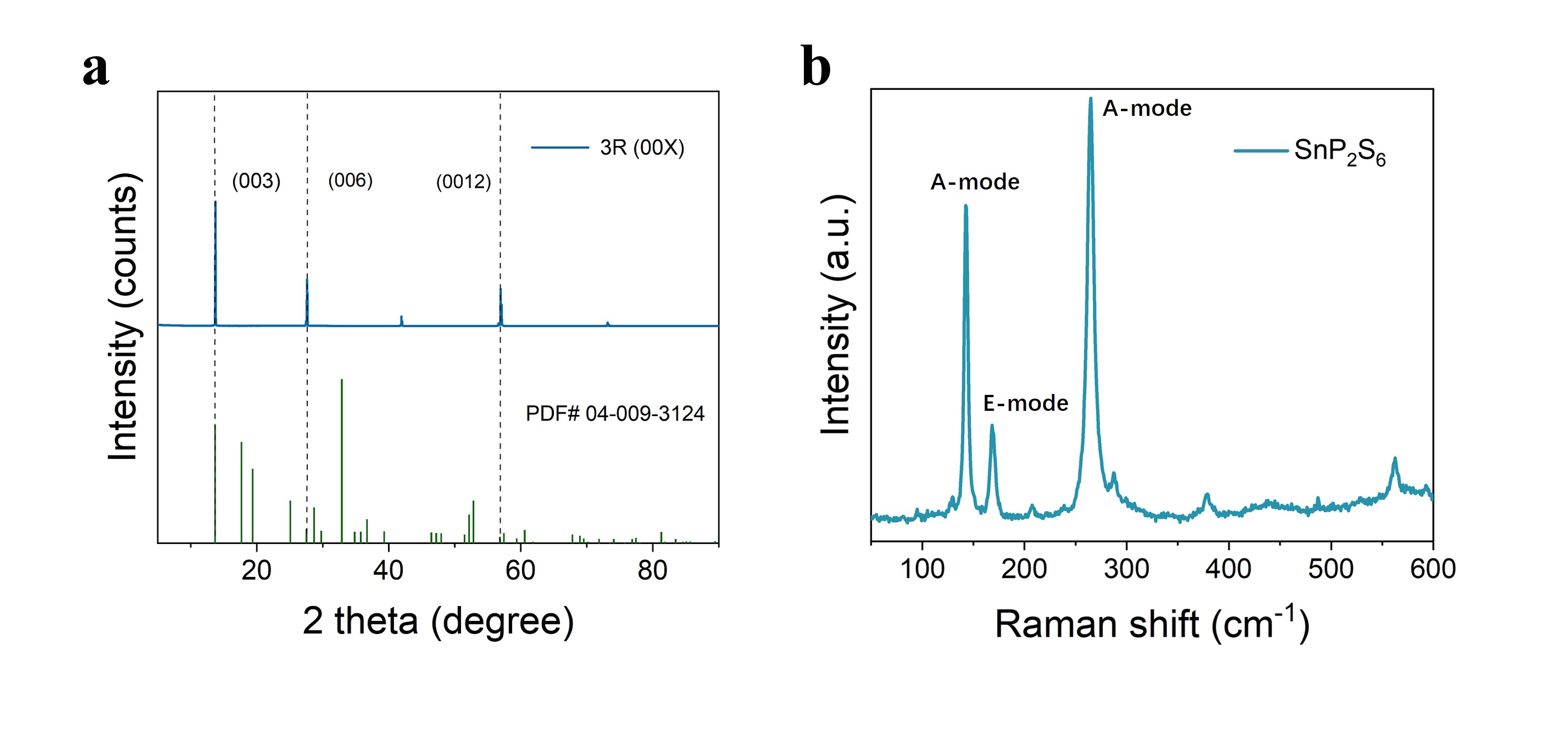}\vspace{-3pt}
	\caption{ Structural characterizations of the synthesized SnP$_2$S$_6$ sample. (a) XRD pattern and (b) Raman spectrum of SnP$_2$S$_6$ singe crystal. }
	\label{Figure S6}
\end{figure}

\begin{figure}[!htb]
	\centering
	\includegraphics[width=1\linewidth]{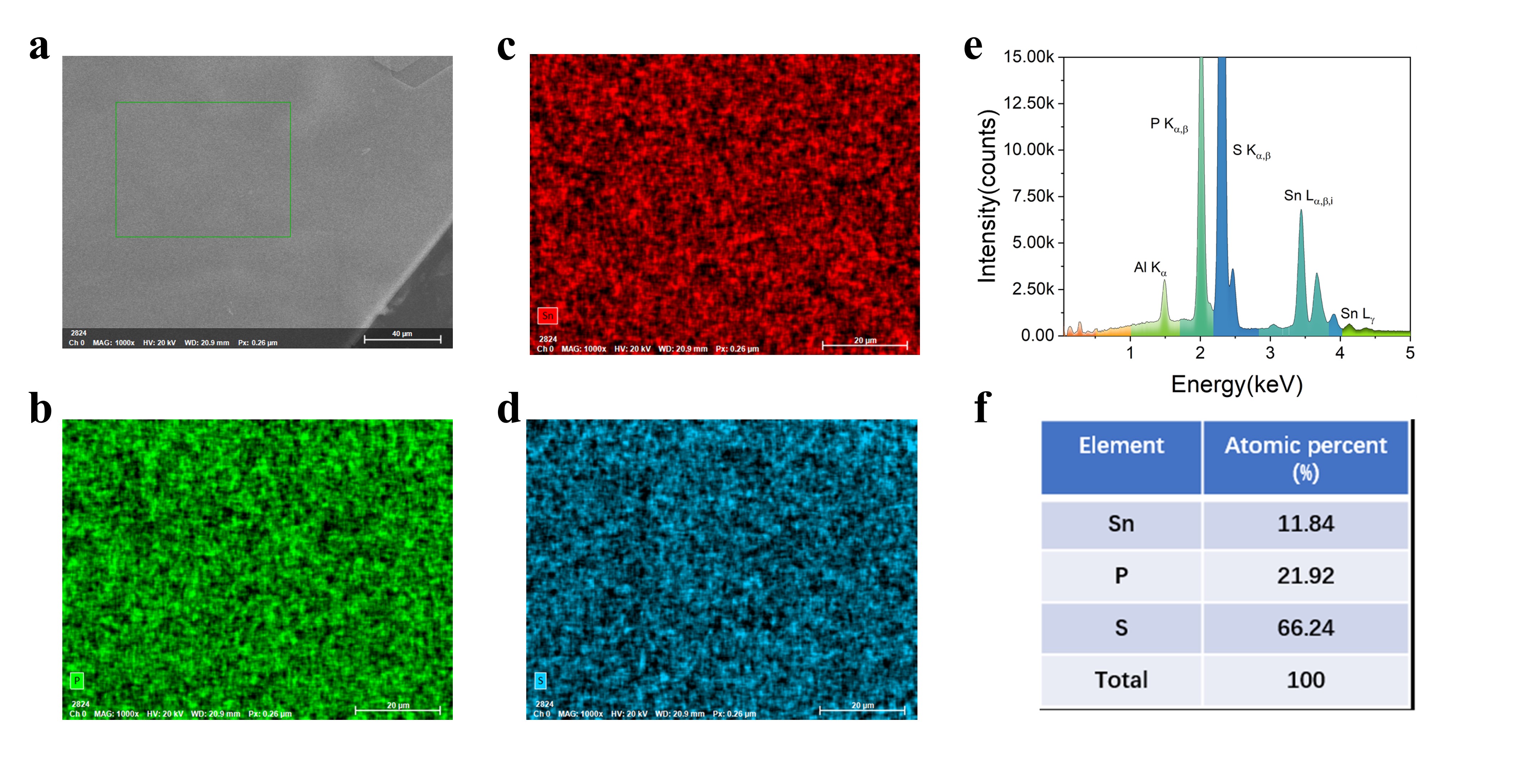}\vspace{-3pt}
	\caption{ (a) SEM image, (b-d) element mapping, and (e-f) quantitative energy dispersive X-ray spectrometer (EDS) spectrum of the single crystal SnP$_2$S$_6$. The uniform distribution of Sn, P and S elements in the SnP$_2$S$_6$ sample is clearly indicated. }
	\label{Figure S7}
\end{figure}

\begin{figure}[!htb]
	\centering
	\includegraphics[width=1\linewidth]{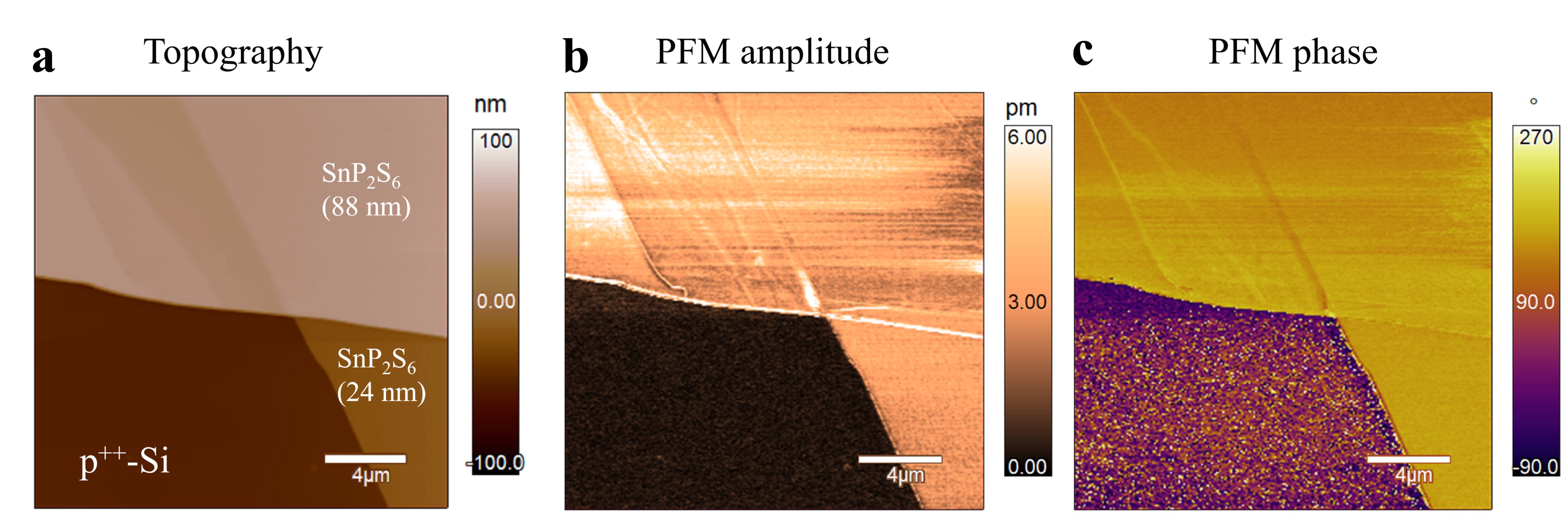}\vspace{-3pt}
	\caption{ The uniform signal in amplitude channel and unicolored phase signal indicate the single-domain state of the SnP$_2$S$_6$ flakes. For a number of SnP$_2$S$_6$ flakes exfoliated from different crystals, majority of them exhibit a monodomain state. (a) Topographic, (b) PFM amplitude, and (c) PFM phase images of exfoliated SnP$_2$S$_6$ flakes on conductive p$^{++}$-Si substrate. }
	\label{Figure S8}
\end{figure}

\begin{figure}[!htb]
	\centering
	\includegraphics[width=1\linewidth]{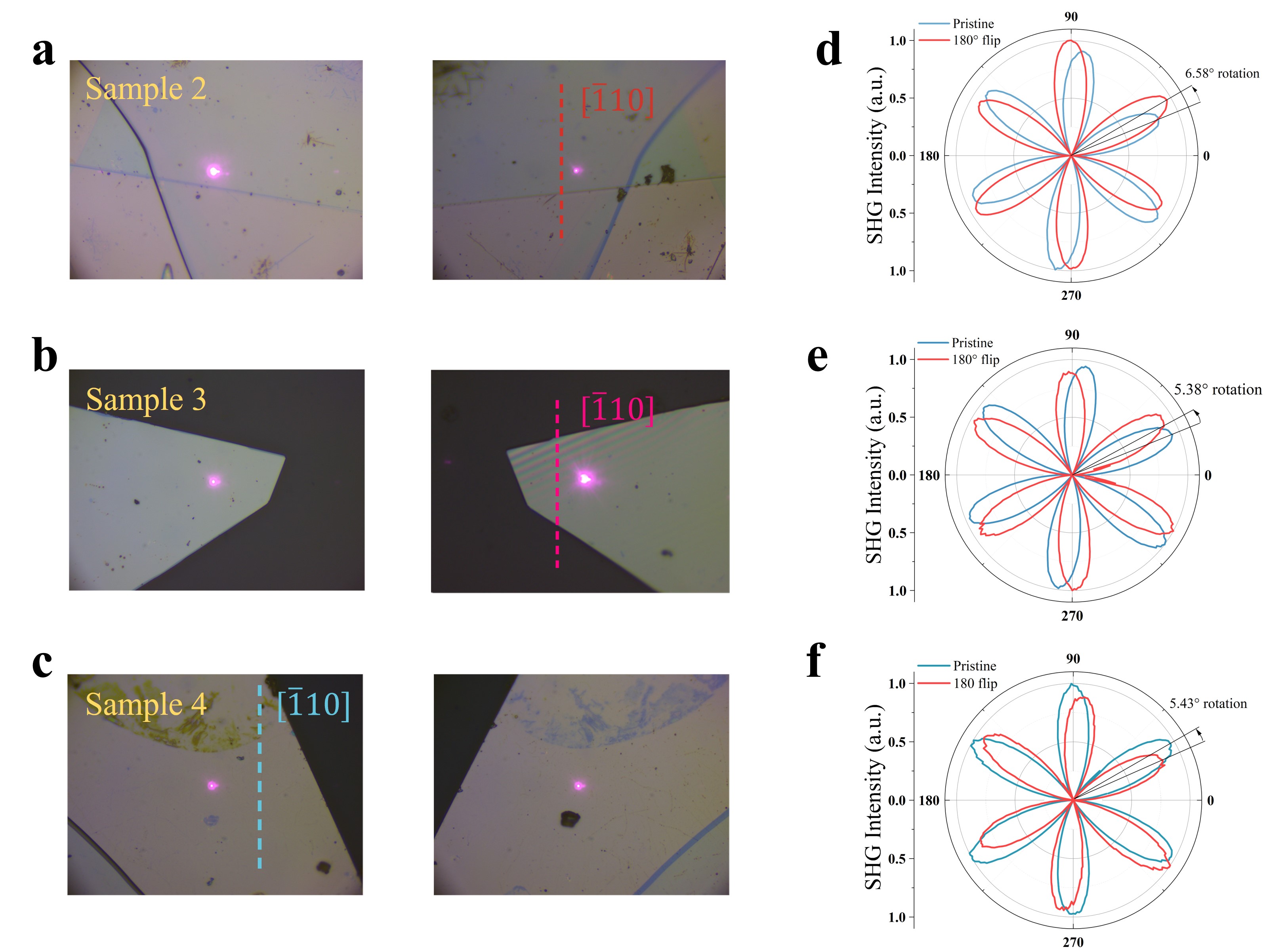}\vspace{-3pt}
	\caption{ (a-c) The optical morphology for several SnP$_2$S$_6$ samples collected before and after flipping their out-of-plane polarizations. To calibrate the in-plane orientation of SnP$_2$S$_6$ samples before and after polarization flipping, $[\overline110]$ crystallographic direction of SnP$_2$S$_6$ is also provided in the morphology photos. (d-f) corresponding SHG polar plots ($I_\parallel$ components only) under the incident 1064 nm laser, collected from different SnP$_2$S$_6$ samples before and after polarization flipping. The notable rotation of SHG polar directions upon polarization flipping are detected in all samples.}
	\label{Figure S9}
\end{figure}